\documentclass[%
reprint,
superscriptaddress,
amsmath,amssymb,
aps, physrev,
prb,
floatfix,
]{revtex4-2}
\usepackage{soul}
\usepackage{graphicx}
\usepackage{dcolumn}
\usepackage{bm}
\usepackage{hyperref}

\usepackage{orcidlink}
\hypersetup{colorlinks=true,linkcolor=blue,citecolor=blue}

\begin{document}
\title{Organic–Inorganic Polaritonics: Linking Frenkel and Wannier–Mott Excitons}
\author{V. G. M. Duarte\,\orcidlink{0009-0009-5836-6084}}
\email{vgmduarte@gmail.com}
\affiliation{Department of Physics, Aeronautics Institute of Technology, 12228-900, São José dos Campos, SP, Brazil}
\affiliation{International Iberian Nanotechnology Laboratory (INL), Av Mestre Jos\'e Veiga, 4715-330 Braga, Portugal}
\author{A. J. Chaves\,\orcidlink{0000-0003-1381-8568}}
\affiliation{Department of Physics, Aeronautics Institute of Technology, 12228-900, São José dos Campos, SP, Brazil}
\affiliation{
 POLIMA---Center for Polariton-driven Light--Matter Interactions, University of Southern Denmark, Campusvej 55, DK-5230 Odense M, Denmark
}
\author{N.~M.~R.~Peres\,\orcidlink{0000-0002-7928-8005}}
\affiliation{International Iberian Nanotechnology Laboratory (INL), Av Mestre Jos\'e Veiga, 4715-330 Braga, Portugal}
\date{\today}
\affiliation{
 POLIMA---Center for Polariton-driven Light--Matter Interactions, University of Southern Denmark, Campusvej 55, DK-5230 Odense M, Denmark
}
\affiliation{Centro de F\'{\i}sica (CF-UM-UP) and Departamento de F\'{\i}sica, Universidade do Minho, P-4710-057 Braga, Portugal}

\begin{abstract}
In recent years, organic materials have emerged as promising candidates for a variety of light-harvesting applications ranging from the infrared to the visible regions of the electromagnetic spectrum. Their enhanced excitonic binding energies and large transition dipole moments enable strong coupling with light, with some systems already reaching the ultrastrong coupling regime. In contrast, a wide range of two-dimensional (2D) materials has been extensively explored in the literature, exhibiting high exciton stability and strong electron–hole coupling due to reduced screening effects.
In this Letter, we present a microscopic model describing the interaction of 2D materials and organic molecular aggregates in an optical cavity. We predict the formation of a hybrid Wannier–Mott–Frenkel exciton-polariton with an enhanced Rabi splitting, exceeding that of the pure organic cavity by several tens of meV.
To elucidate this phenomenon, we examine a cavity with 2D tungsten sulfide and a cyanine dye, where this enhancement corresponds to a $5\%$ increase relative to the organic cavity. The complementary characteristics of Wannier–Mott and Frenkel excitons enable the formation of tunable polariton states that merge into a single hybrid state as a function of detuning, allowing for dual Rabi splitting mechanisms.
This provides a promising platform for exploring quantum optical phenomena in both the strong and ultrastrong coupling regimes.
\end{abstract}

\maketitle

\begin{figure*}[ht]
    \includegraphics[width=2\columnwidth]{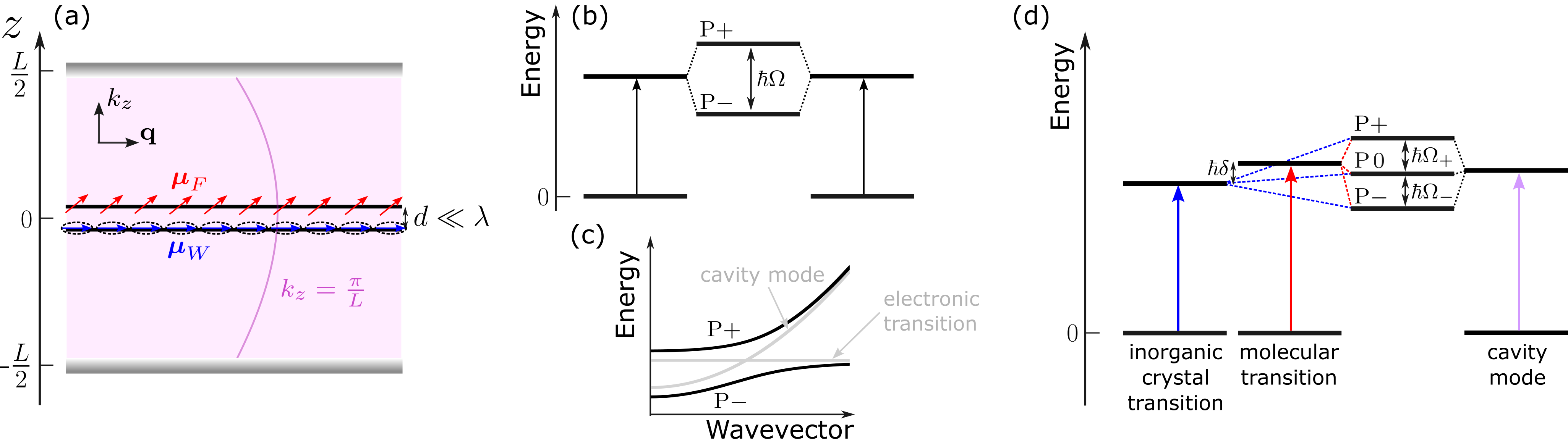}
    \caption{\label{fig:schematic}Light--matter strong coupling in an organic--inorganic Fabry--Pérot cavity.
    (a) Cavity configuration. A two--dimensional (2D) organic material is placed at $z=d/2$ (for instance, a J--aggregated dye), a 2D inorganic semiconductor at $z=-d/2$ (for instance, a transition metal dichalcogenide monolayer), and two optical mirrors at $z=\pm L/2$.
    The materials are coupled to cavity modes, quantized by the wavenumber in the out--of--plane direction $k_z$.
    The in--plave wavevector $\mathbf{q}$ is also indicated.
    The molecules of the organic material have fixed electric dipoles with an out--of--plane shift (here defined as a shift towards the $z$--axis direction), and therefore has an anisotropic optical response.
    Conversely, the electronic transitions of the inorganic material form in--plane confined dipoles that rotate in a circular fashion, generating an isotropic optical response.
    The electric dipole moments $\boldsymbol{\mu}_F,\boldsymbol{\mu}_W$ are indicated.
    (b) Energy level structure of the electronic transition in a material strongly coupled to an optical cavity mode close in energy, generating pairs of polariton states $P_+,P_-$ separated by the Rabi splitting energy $\hbar\Omega$.
    (c) Diagram of the dispersion relation of a polariton, showing the anticrossing pattern around the resonance between the cavity mode and the electronic transition.
    (d) Energy level structure of the electronic transitions of the inorganic and organic materials with an energy detuning $\hbar\delta$, simultaneously coupled to an optical cavity mode close in energy, generating three polariton states $P_+,P_0,P_-$ separated by Rabi splittings $\hbar\Omega_+,\hbar\Omega_-$. The states $P_{\pm}$ are denominated the edge states, and $P_0$, the middle state.}
\end{figure*}

\emph{Introduction.} Strong light–matter interactions have emerged as a well-established field of research, driving advances in light harvesting and quantum optics and enabling the exploration of strong and ultrastrong coupling regimes~\cite{nature_ultrastrong-coupling,nanophot_strong-light-matter,revmodphys_ultrastrong-coupling,acsphot_exploring-lightmatter,science_manipulating-matter}.
Strong coupling is based on the formation of polaritons, quantum states that arise from the coherent energy exchange between photons and excited electronic states, at a faster rate than the decay of either constituent.
In the strong coupling regime, photons are exchanged backwards and forwards between a material and a resonant optical cavity mode---the so--called Rabi oscillations---with frequency $\Omega$.
This phenomenon gives rise to new superposed eigenstates separated by the Rabi splitting $\hbar\Omega$ [Fig.~\ref{fig:schematic}(b)], characterized by an anticrossing pattern on the polariton dispersion [Fig.~\ref{fig:schematic}(c)].
Polaritons are the foundation of a plethora of groundbreaking discoveries, ranging from lasing and Bose--Einstein condensation~\cite{nature_exciton-polariton-condensates,revmodphys_exciton-polariton-bose-einstein} to quantum memories~\cite{PhysRevA.65.022314}.

Among the polariton--hosting materials, organic materials stand out as a promising platform due to their excitonic states with increased binding energies and large dipole moments.
Organic cavities have shown great potential for polaritonics, presenting large Rabi splittings up to several hundreds of meV, placing them in the ultrastrong coupling regime~\cite{Orgiu2015,PhysRevLett.106.196405}.
Moreover, organic molecule--cavity systems have also been examined from the point of view of molecular vibrational spectra~\cite{Shalabney2015,PhysRevX.5.041022} and polariton dynamics~\cite{Schwartz_polariton-dynamics}, demonstrating long polariton lifetimes and potential for molecule--cavity coupling at room temperature.

Some molecular materials can undergo a reversible process of aggregation, characterized by sharp absorption bands with a dramatic shift in wavelength~\cite{jelley1936,jelley1937,scheibe1936,scheibe1937,scheibe1938}.
In particular, J--aggregates---assemblies of organic molecules arranged in a head--to--tail fashion---garnered significant attention as a result of their remarkable optical properties from the near infrared to the visible range, such as sharp absorption bands and high oscillator strengths~\cite{Jagg_review}.
The unique optical properties of J--aggregates stem from coherent interactions among closely packed molecule assemblies, which result in enhanced dipole--dipole coupling and a delocalization of the excited molecular states.
These excited states behave as Frenkel excitons~\cite{Kasha1,Kasha2}, and are the foundation of modern quantum optics applications stemming from efficient light absorbers~\cite{Jagg_solarcells1,Jagg_solarcells2} to biological imaging~\cite{Jagg_bioimaging}.
The potential of J--aggregates for optical cavity design has been demonstrated through a cyanine J--aggregate, where Rabi splittings up to 500 meV have been observed~\cite{Wang2014}.
Moreover, hybrid systems of organic and inorganic materials have already been proposed in the literature, using molecular crystals and semiconductor quantum wells~\cite{Rocca}, metal--organic interfaces coupled via plexciton resonances~\cite{acsphotonics.4c01219}, and even inorganic--metal--organic setups~\cite{TDBC-WS2_theory,TDBC-WS2_experiment}.
However, these hybrid cavities still offer vast potential for exploration.

In this Letter, we present a microscopic theory for the hybrid organic--inorganic microcavity, where the material's charge excitations interact indirectly via cavity photons.
For enhanced stability and electron--hole coupling, two--dimensional (2D) materials are considered.
Using this framework, we derive a $3 \times 3$ Hamiltonian on the basis of cavity photons, Wannier--Mott excitons, and  Frenkel excitons.
Through this methodology, we investigate cavity polaritons, examining their hybridization and strong coupling compared to the pure organic and inorganic cavities.
By exploring the behavior of this intertwined system, we describe its most important properties for the context of strong coupling in cavity quantum electrodynamics and unveil their potential advantages and unique tunability opportunities.
Additional details on the formalism underpinning this work are provided in the Supplemental Material (SM) accompanying the Letter.

\emph{Resonant exciton--photon interactions.} We consider the configuration presented in Fig.~\ref{fig:schematic}(a).
An organic material and an inorganic semiconductor are positioned close to the $z=0$, where the field amplitude is maximized.
The interlayer distance $d\ll \lambda$---where $\lambda$ is the wavelength of a photon in the near infrared--visible range---is assumed to exceed the range of direct electrostatic interactions between the materials.
The semiconductor serves as a platform for Wannier--Mott excitons, while the molecular material supports Frenkel excitons.
The total Hamiltonian of this system is (see SM~\cite{SM})
\begin{eqnarray}
    \hat{H} = \hbar&&\sum_{\mathbf{q}} \bigg\{\omega_F \hat{B}_{\mathbf{q}}^\dagger \hat{B}_{\mathbf{q}} + \sum_{\lambda} \bigg[\omega_q^{ph} \hat{b}_{\mathbf{q}\lambda}^\dagger \hat{b}_{\mathbf{q}\lambda} + \omega_W \hat{C}_{\mathbf{q}\lambda}^\dagger \hat{C}_{\mathbf{q}\lambda}\bigg]\bigg\} \nonumber\\
    &&
    + \hbar\sum_{\mathbf q \lambda} \bigg\{f_{\mathbf{q}\lambda} \hat{B}_{\mathbf{q}}^\dagger \hat{b}_{\mathbf q \lambda}
    + g_{\mathbf{q}\lambda} \hat{C}_{\mathbf{q}\lambda}^\dagger \hat{b}_{\mathbf{q}\lambda} + \text{H.c.}\bigg\}, \label{eq:H}
\end{eqnarray}
where $\hat{b}_{\mathbf{q}\lambda},\hat{B}_{\mathbf q},\hat{C}_{\mathbf q \lambda}$ ($\hat{b}_{\mathbf{q}\lambda}^\dagger,\hat{B}_{\mathbf q}^\dagger,\hat{C}_{\mathbf q \lambda}^\dagger$) are the boson annihilation (creation) operators of cavity photons, Frenkel excitons and Wannier--Mott excitons with in--plane wavevector $\mathbf{q}$, $\omega_q^{ph}=c\sqrt{q^2+(\pi/L)^2}$ is the bare cavity dispersion, $\omega_F,\omega_W$ are the exciton energies, and $f_{\mathbf q \lambda}, g_{\mathbf q \lambda}$ are dipolar coupling functions of excitons with the cavity electromagnetic field:
\begin{subequations}\label{eq:fg}
    \begin{eqnarray}
        f_{\mathbf{q}\lambda} \hspace{-0.1cm}&=& \hspace{-0.08cm}2\bar{\mu}_F\sqrt{\alpha }\bigg(\hspace{-0.1cm}\sqrt{\omega_c\omega_q^{ph}} \hspace{-0.05cm}\sin\phi\delta_{\lambda 1} \hspace{-0.1cm}+\hspace{-0.05cm} i\sqrt{\frac{\omega_c^3}{\omega_{q}^{ph}}}\hspace{-0.05cm}\cos\phi\delta_{\lambda 2}\hspace{-0.1cm}\bigg)\hspace{-0.05cm}, \label{eq:f}
        \\
        g_{\mathbf{q}\lambda} \hspace{-0.1cm}&=&\hspace{-0.08cm} 2\bar{\mu}_W \sqrt{\alpha} \hspace{-0.1cm}\left(\hspace{-0.1cm}-\sqrt{\omega_c\omega_q^{ph}} \delta_{\lambda 1} \hspace{-0.1cm}+ \hspace{-0.1cm}\sqrt{\frac{\omega_c^3}{\omega_q^{ph}}} \delta_{\lambda 2}\hspace{-0.1cm}\right)e^{i\phi}, \label{eq:g}
    \end{eqnarray}
\end{subequations}
where $\bar{\mu}_F=\mu_F/e$, $\bar{\mu}_W=\mu_W/e$ are the dimensionless in--plane electric dipole moments, $\omega_c=c\pi/L$ is the fundamental cavity frequency, $\lambda=1,2$ represents the photon polarization [$\lambda=1$ for transverse electric (TE) and $\lambda=2$ for transverse magnetic (TM)] as well as the optically active Wannier--Mott exciton modes ($\lambda=1$ for transverse and $\lambda=2$ for longitudinal, relative to $\mathbf{q}$), $\alpha=e^2/(4\pi\epsilon_0\hbar c)$ is the fine structure constant, and $\phi$ is the angle between $\mathbf{q}$ and the in--plane projection of the Frenkel exciton dipole moments.

The coupling functions in Eqs.~\eqref{eq:fg} reveal a clear contrast between Frenkel and Wannier--Mott excitons.
While Wannier--Mott excitons~\eqref{eq:g} exhibit isotropy on the direction of $\mathbf{q}$, with a dependency on $\phi$ given by a phase factor $e^{i\phi}$, Frenkel excitons~\eqref{eq:f} display optically active modes that strongly depend on $\phi$.
For Frenkel excitons, the coupling with TE(TM)--polarized light is maximized when the light propagates perpendicular (parallel) to the dipoles.
In contrast, both coupling functions exhibit a $\mathbf{q}$-dependence of the form $\omega_q^{ph} / \omega_c = \sqrt{(qL / \pi)^2 + 1}$.
For long wavelengths, where $q \ll \pi / L$, we have $\omega_q^{ph}/\omega_c \approx 1$.
Furthermore, if we restrict the coupling functions to the angles $\phi$ where maximum coupling with light is achieved, the polarization dependence of the coupling functions reduces to phase factors $\pm 1$ or $\pm i$.
Since our primary focus is on the strong coupling regime and polariton hybridization within the cavity, and because these polarization dependent phase factors of the coupling functions do not influence the polariton dispersions and state compositions, we omit the index $\lambda$ in subsequent discussions.
In this picture, we have
\begin{equation}
    \frac{|f_{\mathbf{q}}|}{\bar{\mu}_F} = \frac{|g_{\mathbf q}|}{\bar{\mu}_W} \approx \frac{\Omega_0}{2},
\end{equation}
where $\Omega_0 = 4\sqrt{\alpha}\omega_c$.
This shows the symmetry between the coupling functions and will also prove useful in establishing the energy scale of Rabi splittings.

Introducing a polariton operator of the form $\hat{A}_{\mathbf{q}r} = P_{\mathbf{q}r}\hat{b}_{\mathbf{q}} + W_{\mathbf{q}r}\hat{C}_{\mathbf{q}} + F_{\mathbf{q}r} \hat{B}_{\mathbf{q}}$, such that the Hamiltonian~\eqref{eq:H} becomes diagonal ($\hat{H} = \hbar\sum_{\mathbf{q}r} \omega_{\mathbf{q}r} \hat{A}_{\mathbf{q}r}^{\dagger} \hat{A}_{\mathbf{q}r}$), we obtain the secular equation
\begin{equation}\label{eq:H1}
    \begin{pmatrix}
        \omega_q^{ph} & g_{\mathbf{q}} & f_{\mathbf{q}} \\
        g_{\mathbf{q}}^{*} & \omega_W & 0 \\
        f_{\mathbf{q}}^{*} & 0 & \omega_F
    \end{pmatrix}
    \begin{pmatrix}
        P_{\mathbf{q}r}\\
        W_{\mathbf{q}r}\\
        F_{\mathbf{q}r}
    \end{pmatrix}
    =
    \omega_{\mathbf{q}r}
    \begin{pmatrix}
        P_{\mathbf{q}r}\\
        W_{\mathbf{q}r}\\
        F_{\mathbf{q}r}
    \end{pmatrix},
\end{equation}
whose solution yields three polariton branches $\omega_{\mathbf{q}r}$ ($r=-1,0,1$).
Note that this $3\times 3$ formulation is valid only under the assumption of maximized light coupling with a single polarization mode TE or TM, as previously discussed.
In the general case, one arrives at a $5 \times 5$ secular equation and $5$ polariton branches with mixed polarizations~\cite{Rocca}.
Nevertheless, the $3 \times 3$ formulation provides greater clarity, is more suitable for our purposes, and can be easily implemented experimentally.
The coefficients $|P_{\mathbf{q}r}|^2$, $|W_{\mathbf{q}r}|^2$, and $|F_{\mathbf{q}r}|^2$ represent the photon, Wannier--Mott and Frenkel exciton fractions of a polariton state, respectively.

\begin{figure}[ht]
	\includegraphics[width=\columnwidth]{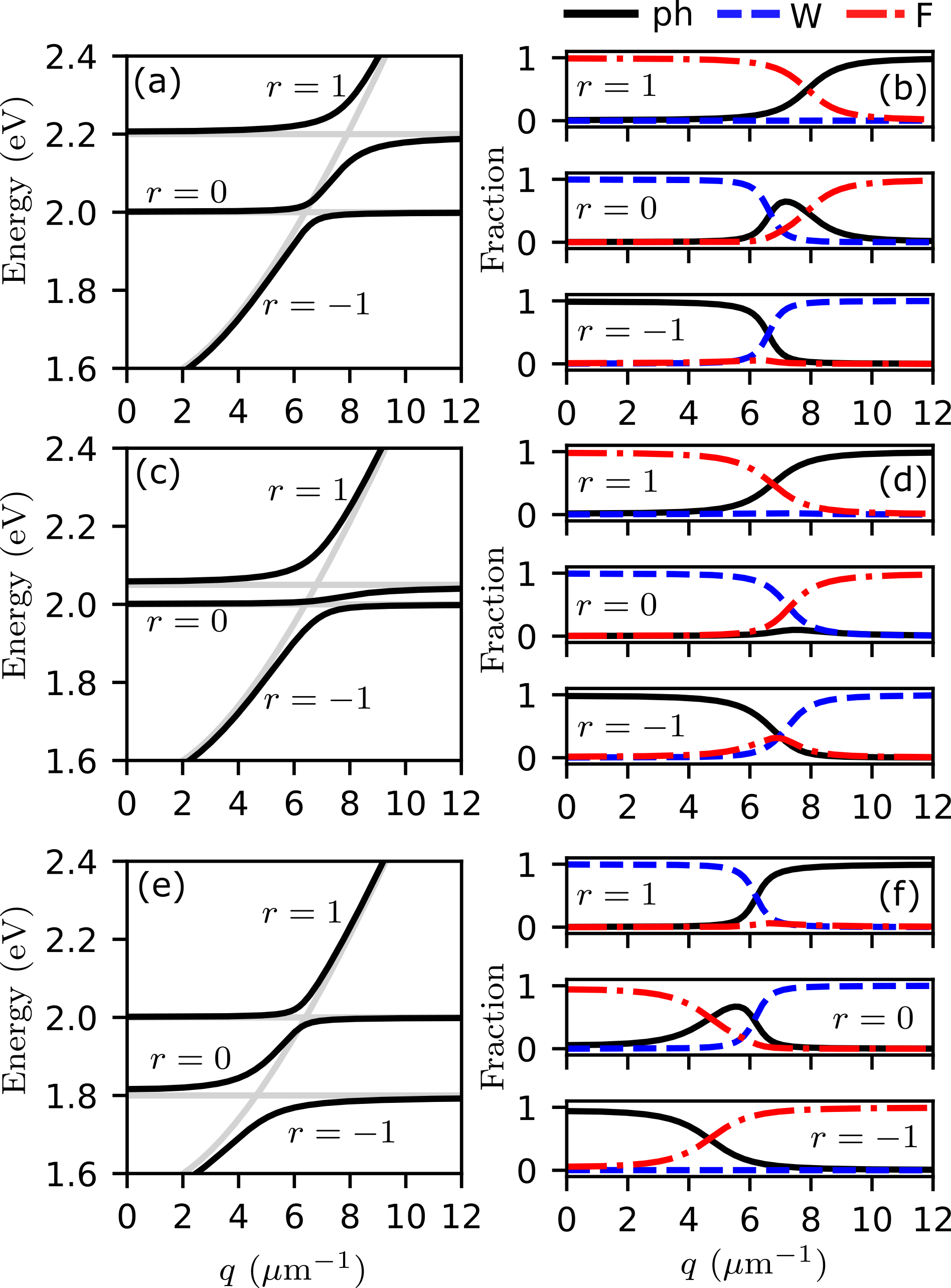}
	\caption{\label{fig:hyb}
    Hybridization of Wannier--Mott and Frenkel polaritons as a function of exciton--exciton detuning $\delta=\omega_F-\omega_W$.
    The fixed parameters are $\hbar\omega_W=2$~eV, $L=400$~nm, $\mu_W=0.10$e, and $\mu_F=0.25$e, and $\hbar\omega_F=2.2$~eV (a,b), $2.05$~eV (c,d), and $1.8$~eV (e,f).
    (a,c,e) Polariton dispersion relations.
    The bare cavity and exciton energies $\hbar\omega_q$, $\hbar\omega_W$ and $\hbar\omega_F$ are presented in gray.
    (b,d,f) Hopfield coefficients, showing the fractions of photons (black solid), Wannier--Mott excitons (blue dashed), and Frenkel excitons (red dot--dashed).}
\end{figure}

\emph{Wannier--Mott--Frenkel exciton polaritons.} The dispersion relation of polaritons in the hybrid cavity is presented in Fig.~\ref{fig:hyb}, together with the correspondent fractions, for three different situations: $\omega_F > \omega_W$, $\omega_F \sim \omega_W$, and $\omega_F < \omega_W$.
One can see the formation of two anticrossings, one for each resonance ($\omega_q=\omega_F$ and $\omega_q=\omega_W$).
Only the TE branches are displayed, as the TE and TM modes exhibit nearly identical behavior within the wavevector range considered.
The upper and lower branches ($r=\pm 1$)---referred to as edge branches--- exhibit a dominant character of either Wannier–Mott or Frenkel polaritons, as illustrated by the Hopfield coefficients in Figs~\ref{fig:hyb}(b), (d), and (f).
The molecular exciton dipoles are typically greater than those of excitons in conventional semiconductors, which explains the enhanced Rabi splitting around the Frenkel exciton resonance and the higher fraction of Frenkel excitons in the Wannier-Mott dominant edge branch, compared to the near--zero fraction of Wannier--Mott excitons in the Frenkel dominant edge branch.
The middle branch ($r=0$) exhibits a dual behavior, blending characteristics of Wannier--Mott and Frenkel polaritons.
In the $q$--range between the exciton--photon resonances, it demonstrates a complete threefold hybridized state of photon, Wannier--Mott exciton, and Frenkel exciton.
However, as the exciton energies approach each other, the photon fraction of the middle branch diminishes to zero, causing the middle branch to evolve into a nearly flat band.
In fact, for $\omega_W = \omega_F = \omega_0$, the solution secular problem~\eqref{eq:H1} is both analytical and short:
\begin{subequations}
\begin{eqnarray}
    \omega_{q0} &=& \omega_0,
    \\
    \omega_{q\pm} &=& \frac{1}{2}(\omega_q^{ph} + \omega_0) \pm \frac{1}{2}\sqrt{(\delta_q^{ph})^2 + (\Omega_H)^2},
\end{eqnarray}
\end{subequations}
where $\delta_{q}^{ph} = \omega_q^{ph} - \omega_0$ is the photon--exciton detuning, and $\Omega_H = \Omega_{0}\sqrt{\bar{\mu}_F^2+\bar{\mu}_W^2}$.
The Hopfield coefficients are
\begin{subequations}
\begin{align}
    &P_{\mathbf{q}0} = 0,\,
    \hspace{-0.3cm}&W_{\mathbf{q}0} = -\frac{2f_{\mathbf{q}}}{\Omega_H},\,
    &F_{\mathbf{q}0} = \frac{2g_{\mathbf{q}}}{\Omega_H},
    \\
    &P_{\mathbf{q}\pm} = \frac{2(\omega_{q\pm}-\omega_0)}{Z_{q\pm}},\,
    \hspace{-0.3cm}&W_{\mathbf{q}\pm} = \frac{2 g_\mathbf{q}^*}{Z_{q\pm}},\,
    &F_{\mathbf{q}\pm} = \frac{2 f_\mathbf{q}^*}{Z_{q\pm}},
\end{align}
\end{subequations}
where $Z_{q\pm} = \sqrt{4(\omega_{q\pm}-\omega_0)^2 + (\Omega_H)^2}$.
These equations clearly illustrate the formation of a single polariton exhibiting hybrid Wannier–Mott and Frenkel characteristics, with a Rabi splitting $\Omega_H$, as depicted in Fig.~\ref{fig:hyb}(c).
As shown in Fig.~\ref{fig:hyb}(d), the photon fractions are proportional to the polariton shift $\omega_{q\pm}-\omega_0$, while the exciton fractions are determined by the exciton–photon coupling functions.
When the photon--exciton detuning is zero, the photon fraction is equal to the combined Wannier--Mott and Frenkel exciton fractions.

\begin{figure}[ht]
    \includegraphics[width=\columnwidth]{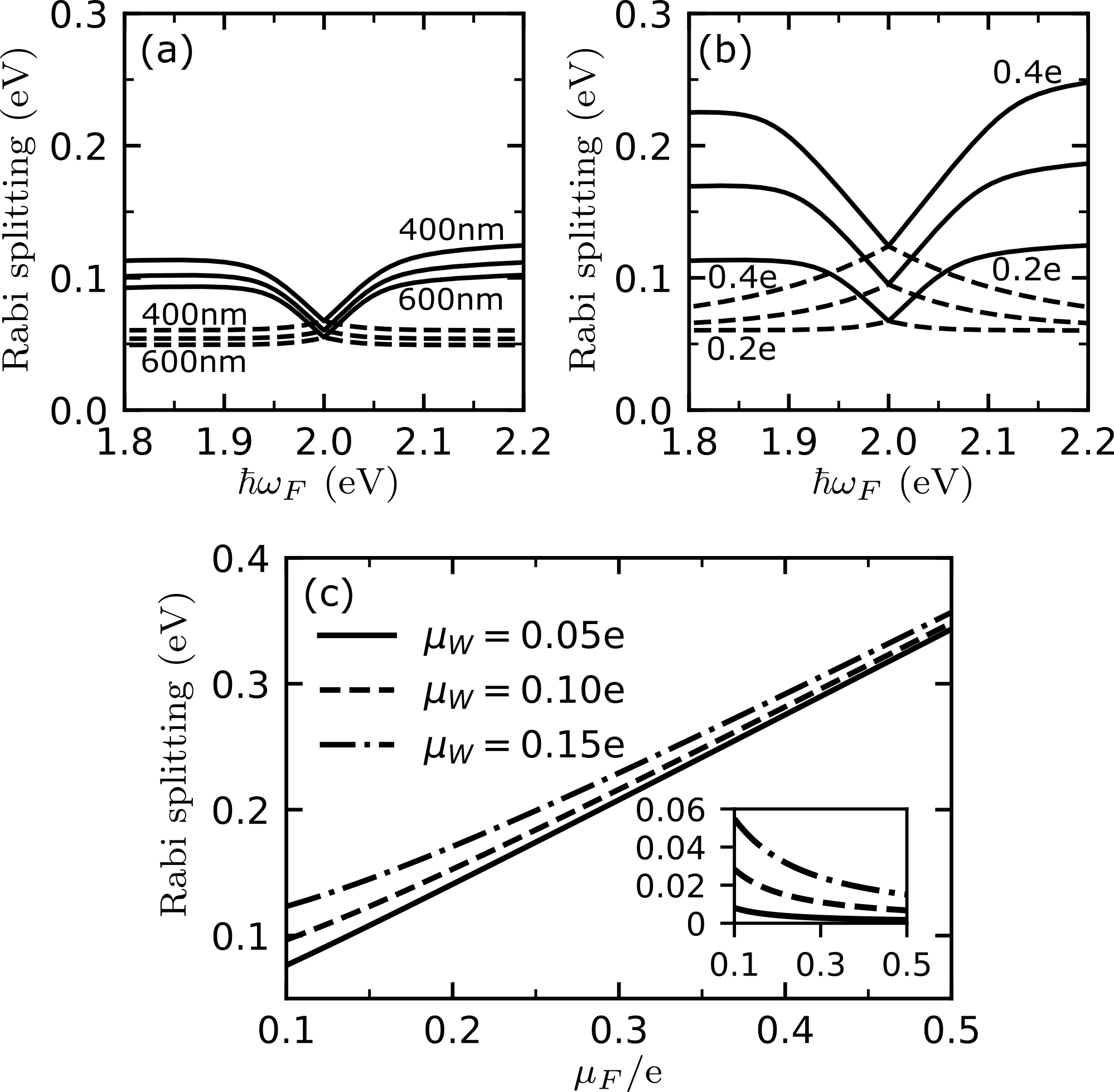}
    \caption{\label{fig:split}Dual Rabi splitting of the hybrid cavity.
    The parameters are, unless stated otherwise, $\hbar\omega_W=\hbar\omega_F=2$~eV, $L=400$~nm, $\mu_W=0.10$e, and $\mu_F=0.2$e.
    (a,b) Evolution of the Rabi splittings $\Omega_F$ (solid) and $\Omega_W$ (dashed) versus Frenkel exciton energy $\hbar\omega_F$, as (a) the cavity width $L$ decreases from $600$nm to $400$nm and (b) the Frenkel exciton dipole $\mu_F$ increases from $0.2$e to $0.4$e.
    (c) Hybrid Rabi splitting $\Omega_H = \Omega_W + \Omega_F$ versus $\bar{\mu}_F=\mu_F/e$, for $L=310$~nm ($\hbar\omega_c=2$~eV) and different values of the Wannier--Mott exciton dipole $\mu_W$.
    Inset: $\Omega_H-\bar{\mu}_F \Omega_{0}$ (eV) versus $\bar{\mu}_F$, where $\bar{\mu}_F \Omega_{0}$ is the pure Frenkel polariton Rabi splitting.}
\end{figure}

We denote the Rabi splitting energy associated to the anticrossing of the $\omega_q=\omega_F$ ($\omega_q=\omega_W$) resonance as $\hbar\Omega_F$ ($\hbar\Omega_W$).
To further examine the hybridization and coupling strength of the Wannier–Mott–Frenkel polariton, Figs.~\ref{fig:split}(a) and (b) show the evolution of $\hbar\Omega_W$ and $\hbar\Omega_F$ as a function of the exciton--exciton detuning $\delta=\omega_F-\omega_W$, demonstrating that the splittings converge to the same value as $\delta$ approaches zero.
In Fig~\ref{fig:split}(a) it is shown that, as $L$ decreases, the splittings get larger.
For a similar reason, in Fig.~\ref{fig:split}(b) it is shown how the splitting evolves as $\mu_F$ is increased, while keeping $\mu_W$ fixed.
In fact, the splittings depend directly on the coupling strength between excitons and photons, mediated by the coupling functions $f_{\mathbf{q}}$ and $g_{\mathbf q}$.
The larger these functions, the larger the splittings will be.
These functions, in turn, depend linearly on the transition dipole magnitudes $\mu_F,\mu_W$, as well as the fundamental cavity frequency $\omega_c$, which in turn depends on $L^{-1}$.
We verified that, for both the panels (a) and (b) of Fig.~\ref{fig:split}, the Rabi splitting $\Omega_F$ ($\Omega_W$) approaches the pure organic (inorganic) cavity Rabi splitting $\bar{\mu}_F \Omega_{0}$ ($\bar{\mu}_W \Omega_{0}$)
at large $\delta$, where the polaritons have low hybridization.
The pure polariton splittings are proportional to the dipole moments of the corresponding excitons, which explains why, in Fig.~\ref{fig:split}(b), the splitting $\Omega_W$ approaches a fixed value as $\mu_F$ changes, while $\Omega_F$ increases significantly.
As $\delta$ approaches zero---illustrated in Fig.~\ref{fig:split} when $\hbar\omega_F$ approaches $\hbar\omega_W = 2.0$~eV---the hybridization between polaritons intensifies, and the Rabi splittings converge toward each other.

To quantify how low the detuning must be for this hybrid regime to occur, we compare it to the pure cavity Rabi splittings.
As discussed previously, the hybridization of the polaritons is triggered when the middle branch flattens and the Rabi splittings merge into a single mode with an enhanced frequency of oscillation.
Defining $\omega_> = \max(\omega_W, \omega_F)$ and $\omega_< = \min(\omega_W, \omega_F)$, it is reasonable to assume that the middle branch enters a flat regime when the combined shifts of the middle branch downward at $\omega_q = \omega_>$ and upward at $\omega_q = \omega_<$ exceed $\delta$.
Since the splittings at large $\delta$ approach the pure cavity splittings, we conclude that the hybrid regime is established when $\delta$ is smaller than the half sum of the pure cavity Rabi splittings, hence $|\delta| \lesssim \Omega_0(\bar{\mu}_F+\bar{\mu}_W)/2$.
We verified that this is indeed the case, as the Rabi splittings deviate significantly from the pure cavity behaviors within this detuning range.
In this region, $\Omega_F$ is diminished, while $\Omega_W$ is augmented.
At the limit $\delta=0$, we reach $\Omega_F=\Omega_W$.
However, at zero detuning, we should consider the combined Rabi splitting $\Omega_H = \Omega_F + \Omega_W = \Omega_0\sqrt{\bar{\mu}_F^2+\bar{\mu}_W^2}$.
Therefore, the hybrid polariton exhibits a Rabi splitting that is twice the value shown by the curves in Figs.~\ref{fig:split}(a) and (b) at $\delta=0$.
This is sufficient to conclude that the hybrid cavity enhances the Rabi splitting of the resulting system compared to the pure cavity splittings, as illustrated in Fig.~\ref{fig:split}(c). In this panel, we show the evolution of $\hbar\Omega_H$ as $\mu_F$ increases, revealing a nearly linear and increasing trend.
The splitting enlarges further as $\mu_W$ increases, moving away from the pure Frenkel polariton line.

In Fig.~\ref{fig:split}(c), we include an inset to illustrate how much the hybrid cavity splittings exceed the larger of the two pure cavity splittings, showing increases up to tens of meV.
To illustrate this effect on a real system, we performed calculations for a cavity containing tungsten disulfide (WS$_2$) and tetrachloro--diethyl--benzimidazolo--carbocyanine (TDBC).
The former is an inorganic crystal with a Wannier--Mott dipole moment $\mu_W=\gamma\beta I$, where $\gamma$ is a parameter of the effective mass Dirac Hamiltonian, $\beta$ is a variational parameter of the exciton wavefunction $\psi_\beta(r)\propto e^{-r/\beta}$, and $I$ is an integral in momentum space (see SM~\cite{SM}).
The ground state exciton of WS$_2$ has an effective mass of $0.16 m_0$---where $m_0$ is the bare electron mass--- and a screening parameter $r_0 \approx 38$~\AA~\cite{PhysRevB.88.045318}, from which it follows that $\beta \approx 13$~\AA~and, since $\gamma = 3.34$~eV~\AA~\cite{TMD_kp_theory}, we get $\bar{\mu}_W \approx 0.1$.
The J--aggregate TDBC hosts Frenkel excitons, and the dipole--dipole coupling between molecules is responsible for a redshift in the optical spectrum.
Since microscopic data on the alignment of the dipoles in the direction of aggregation, as well as on the out--of--plane shift, is still lacking in the literature, we must perform some approximations.
Using the usual Hamiltonian of molecular excitons in the Heitler--London approximation, we derive the shorthand rule
\begin{equation}
    \bar{\mu}_F \approx \sqrt{\frac{a \Delta_J}{6.9\,\text{eV nm}}},
\end{equation}
where $\Delta_J$ (eV) is the redshift and $a$ is the molecule size.
For TDBC, we have $a \approx 2$~nm~\cite{Struganova2002} and $\hbar\omega_0-E_F \approx 0.3$~eV~\cite{Jagg_review}, which yields $\bar{\mu}_F \approx 0.3$.
Therefore, with the exciton energies set at $\hbar\omega_W \approx \hbar\omega_F \approx 2$~eV and the cavity photon energy at $\hbar\omega_c = 2$~eV, we obtain $\hbar\bar{\mu}_F\Omega_0 = 205$~meV and $\hbar\Omega_H = 216$~meV.
This indicates an increase in the Rabi splitting by $11$~meV, corresponding to a $5.4\%$ enhancement.
Given that typical exciton and cavity photon decay rates are reported to be of the order of a few meV~\cite{PhysRevLett.123.067401,PhysRevX.7.021026} and $10^{-3}$ meV~\cite{Takahashi:07}, respectively, these Rabi splittings exceed the decay rates of their constituents by at least one to two orders of magnitude, confirming the strong coupling regime.

\emph{Conclusions.}
Using a dipolar coupling model, we presented the first microscopic description of light–matter coupling in a cavity containing both 2D organic and inorganic materials.
The optical resonator enhances the exciton--photon interaction, leading to quantum states that alternate between Wannier--Mott and Frenkel exciton polariton behavior.
The key differences in the dipole moments of each type of exciton were emphasized.
The formation of a hybrid Wannier--Mott--Frenkel polariton was discussed, with two characteristic Rabi--splittings whose hybridization is enhanced by decreased exciton energy detuning.
Our model is applicable to a wide range of combinations of organic materials and conventional 2D semiconductors.
The hybridization leads to an enhanced Rabi splitting expressed as $\Omega_H = \Omega_0\sqrt{\bar{\mu}_F^2 + \bar{\mu}_W^2}$, effectively merging the two polaritons with individual Rabi splittings $\bar{\mu}_F\Omega_0$ and $\bar{\mu}_W\Omega_0$.
For real systems, this enhancement leads to an increase of tens of meV relative to the organic material, which already exhibits higher Rabi splittings than the inorganic counterpart due to its stronger dipole moments.
By carefully selecting material pairs, this approach could potentially extend into the ultrastrong coupling regime.
The model can be easily extended to account for complexities such as additional layers, excitonic energy spectra, organic materials with more than one type of molecule or dipoles with alternating orientations.

\emph{Acknowledgments.}
V.~G.~M.~D. acknowledges a PhD. Scholarship from the Brazilian agency Capes (Fundação Coordenação de Aperfeiçoamento de Pessoal de Nível Superior).
N.~M.~R.~P. acknowledges support from the European Union
through the EIC PATHFINDER OPEN project No.
101129661-ADAPTATION, and the Portuguese Foundation for Science and Technology (FCT) in the framework
of the Strategic Funding UIDB/04650/2020, COMPETE
2020, PORTUGAL 2020, FEDER, and through project
PTDC/FIS-MAC/2045/2021. A.~J.~C. acknowledges  CNPq (Conselho Nacional de Desenvolvimento Cient\'ifico e Tecnol\'ogico) Grant No. 423423/2021-5, 408144/2022-0, and 315408/2021-9.

\bibliography{references}
\end{document}